\documentclass[aps,prl,preprint,amsfonts,amsmath,amssymb,a4paper,final,showkeys,showpacs,endfloats*]{revtex4-1}

\usepackage{graphicx}
\usepackage{graphics}
\usepackage{dcolumn}
\usepackage{bm}
\usepackage{epsfig}
\usepackage{multirow}
\usepackage{longtable}
\usepackage{rotating}
\usepackage[pdftex]{hyperref}
\usepackage[all]{hypcap}
\hypersetup{
    backref=True,
    pagebackref=True,
    pdffitwindow=True,     
    pdfstartview=FitH,    
    pdfpagemode=UseNone,
    pdfview=FitH,
    colorlinks=True,        
    linkcolor=blue,         
    citecolor=blue,         
    filecolor=blue,         
    urlcolor=blue           
}

\begin{document}
\setcounter{page}{1}

\title{Microscopic Property of Amorphous Semiconductor Metal Oxide InGaZnO$_{4}$ and Role of O-deficiency}
\author{Il-Joon \surname{Kang}$^{1,2}$ and C. H. \surname{Park}$^{1,2,3}$}

\affiliation{$^1$Research Center for Dielectric and Advanced Matter Physics}
\affiliation{$^2$Department of Physics}
\affiliation{$^3$Department of Physics Education, Pusan National University, Busan 609-735, Republic of Korea}

\begin{abstract}
We investigated the microscopic and electronic structures amorphous oxide semiconductors InGaZnO$_{4}$ (a-IGZO) and the role of O-deficiency through the first-principle calculations. The structure of the amorphous oxide is complicated by the admixture of many different kinds of substructures, however it is surprisingly found that the band tail states, which are well-known to be present in the amorphous semiconductors, are few generated for the conduction band minimum (CBM). The electronic structure around CBM is little affected by the disorder and also by the O-deficiency. Free electron carriers can be generated without a creation of donor-level in the O-deficient amorphous oxide. 
\end{abstract}

\maketitle

Recently the amorphous semiconductor metal-oxides (ASMO) such as InGaZnO (a-IGZO), ZnSnO, and InSnO(ITO)
\cite{nomura1,kamiya2,jkyeong} are extensively investigated, since they can be base materials for the thin-film transistor (TFT) of flat panel display as well as transparent conducting oxides (TCO). The n-type conductivity of ASMOs can be much higher than that of the conventional semiconductors such as as a-Si:H, since the conduction path of electron carriers is characterized mainly by non-directional spherical cation-\emph{s} orbitals of the metal elements \cite{nomura2}. The O-deficiency is suggested to be a important source of high electrical conductivity of ASMO and TCO. The carrier concentration in ASMO can be controlled over a wide range by the O-deficiency\cite{jianke}. However the microscopic understanding of the role of the O-deficiency is not yet resolved with many unresolved aspects. The optical measurements indicate that the O-deficiency induces deep levels in ASMO\cite{kamiya2}, and the first-principles calculations also showed that the O-vacancies in a-IGZO can create various deep or shallow levels\cite{kamiya}. It is suggested that the deep trap centers by the O-deficiency adversely affects the bias and illumination stress instability of oxide TFT \cite{jkyeong,ko}. A puzzling aspect is that the carrier mobility is measured to be much higher in the disordered a-InGaZnO (a-IGZO) than in the well-ordered crystalline IGZO (c-IGZO), unless the O-deficiency is too serious\cite{nomura2}. This is not understood by a simple defect picture of O-vacancy, since the cation-driven dangling bonds around O-vacancies can generate localized levels which can play a role of trap center. Since the microscopic understanding for the O-deficiency is limited, the results from the studies of the O-vacancy in ZnO or other crystalline oxides have been applied to understand the role of O-deficiency. Therefore, the systematic study about the role of the O-deficiency in the ASMO is needed.

In this Letter, we report on the results of first-principles calculations carried out to understand the role of the O-deficiency in a-IGZO as a representative of the ASMO. Our approach is different from a previous study in that we investigate the properties of an O-deficient a-IGZO structures that are generated via the optimization through a molecular dynamics simulated annealing processes. We find that (i) the disordered a-IGZO is mixed by many different kinds of building blocks, however (ii) the band tail states, which is well-known to be present in amorphous materials, can be surprisingly absent for CBM in the amorphous-IGZO, and (iii) the localization character of the CBM is little affected by the O-deficiency in a-oxide, and no localized donor-center is formed, that is fundamentally in contrast to the O-vacancy in crystalline IGZO (c-IGZO), and (iii) the result of the disorder or O-deficiency is to enhance the formation of localized states around the valence band. We propose that these results provide a new perspective to understand the role of O-deficiency in ASMO.

The first-principles calculations were performed by the projector augmented wave (PAW) method \cite{paw} of the Vienna \emph{ab-initio} simulation package (VASP) \cite{vasp}. The Perdew-Burke-Ernzerhof exchange-correlation functional (PBE) \cite{pbe} approach utilizing the generalized gradient approximation (GGA) scheme was employed and the LDA+U method was used to describe the localized semi-core states of Zn-3\emph{d} orbitals \cite{ldau}. These semi-core orbitals are not accurately described by normal LDA calculation, in which their overlap with the O-\emph{p} orbitals is enhanced due to a strong self-interaction effect in the localized orbitals. The strong Coulomb energy at the localized \emph{d} orbitals can be compensated by the LDA+U method. We used U=5 eV for only the Zn-3\emph{d} orbitals. As an accuracy test, the lattice constants of the binary oxides, In$_2$O$_3$, Ga$_2$O$_3$, and ZnO, are calculated to be 12.42 \AA, 10.117 \AA, and 3.283 \AA, respectively, in a good agreement with the experimental values of 12.23 \AA, 10.117 \AA, and 3.25 \AA.

In order to generate a reliable structure for the stoichiometric and O-deficient a-IGZO, we repeated the simulated annealing (SA) steps using molecular dynamic calculations on supercells. The simulated temperature was increased up to 4000 K, and decreased slowly to 0 K and the SA was repeated between 1000 K and 0 K until the total energy and the atomic structure were converged. We tested several supercells containing 56, 84, and 112 atoms. Calculated results on 112 atoms supercell are given here, since the cell size dependence is weak between 84 and 112 atoms cells. The total energy of the a-IGZO generated in this manner is higher by only 0.18 eV/atom than that of c-IGZO.

\begin{figure}
\begin{center}
\label{eff}
\centering{\includegraphics[width=16.5cm]{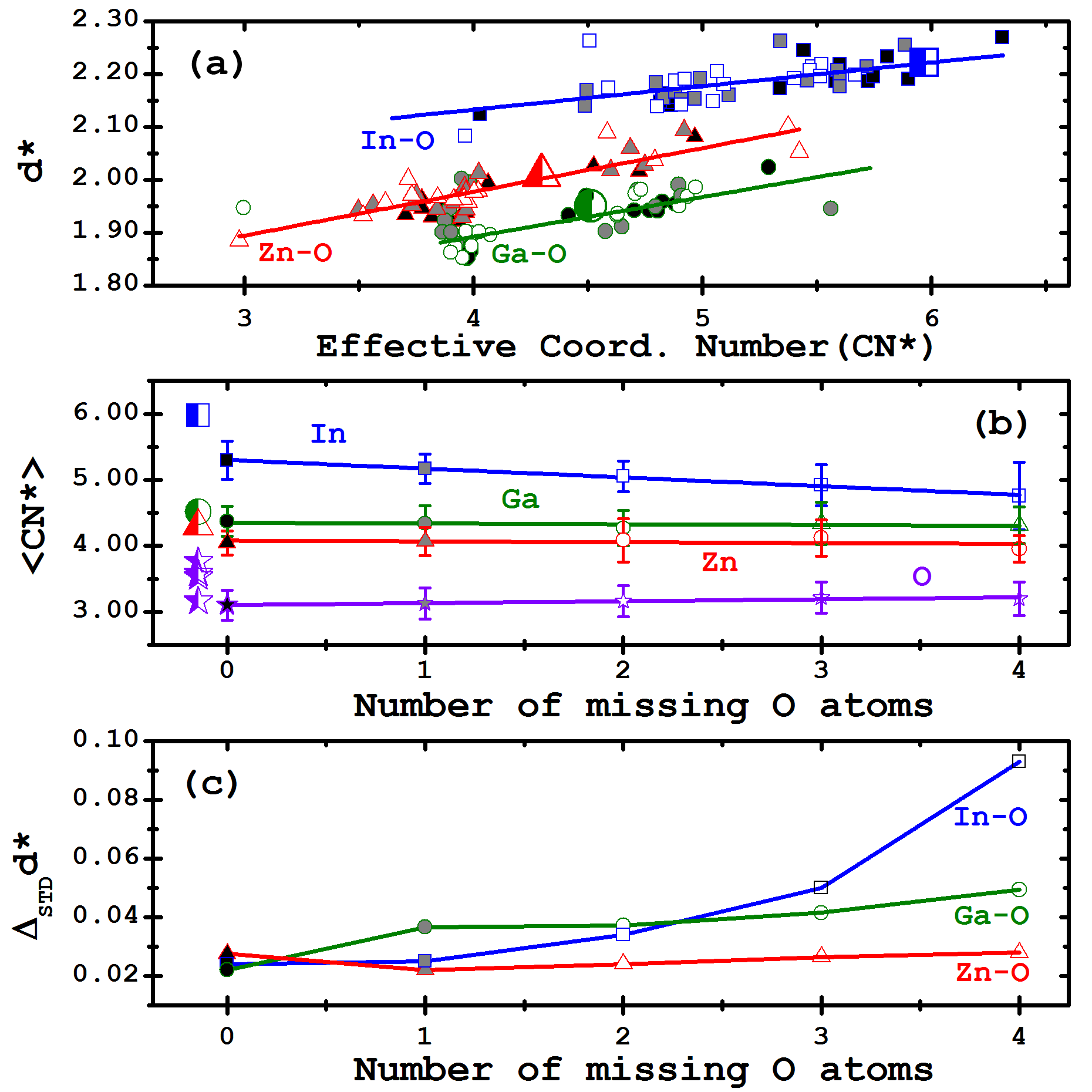}}
\caption{(Color online) (a) The effective coordination number (CN$^*$) versus the effective bond length(d$^*$) in
the stoichiometric a-IGZO are shown by solid symbols. The O-deficient a-IGZO with one and two missing oxygen atoms
are also shown, respectively, by gray and white symbols. The fitting lines are obtained by the importance
samplings. The large half-filled symbols represent the corresponding values in c-IGZO. (b) The average of the
coordination numbers ($<$CN$^*>$) versus $x$, with the standard deviations ($\Delta_{\mathrm{STD}}$CN$^*$)
indicated by the error bar, and (c) the standard deviations of the calculated d$^*$ from the fitting lines
($\Delta_{\mathrm{STD}}$ d$^*$) are shown.}
\end{center}
\end{figure}

In order to understand the microscopic structures of a-IGZO, we examined the short-range orders by calculating the effective coordination numbers (CN$^*$) of cations and the averaged effective bond-length
(d$^*$)\cite{effcn1,effcn2}. We find that {\it the structure of a-IGZO is complicated by the admixture of many kinds of C-O$_{\mathrm{CN}}$ molecular building blocks}, where C denotes a cation and CN is
the coordination number of a cation, as shown in Fig. \ref{eff}a. The nondirectional cation-$s$-obitals make these various substructures accommodated in the ASMO. The average of d$^*$ increases as we go from Ga to Zn to In. In addition, d$^*$ increases with increasing CN$^*$, and this is due to the increased repulsive interaction between O atoms. The graph of d$^*$ vs. CN$^*$  can be nicely fit by a straight line. Fig. \ref{eff}b show that the average ($<$CN$^*>$) and the standard deviation of CN$^*$ ($\Delta_{\mathrm{STD}}$CN$^*$) increases, as we go from Zn to Ga and to In. The In atoms are noted to incorporate the most diverse set of substructures, while Zn prefers mainly a tetrahedral geometry. These indicate that the substructure of In is flexible. Although In and Ga atoms are isoelectric, the In-$<$CN$^*>$ is larger than Ga-$<$CN$^*>$, due to the In¡¯s larger atomic size. The column-III cations donating three electrons can attract more O$^{2-}$ ions than the column-II Zn which donates two electrons. The overall ($<$CN$^*>$) are smaller in a-IGZO than in the c-IGZO. Thus the volume of a-IGZO is larger by 0.6 \%.

Before we examine the O-deficient state of a-IGZO, we investigated the properties of an metastable O-vacancy structures, that are formed by the removal of an O atom from the stoichiometric structure of a-IGZO. The atomic positions were relaxed through the minimization of Hellman-Feynman forces. This case has been tested previously by Kamiya et al.\cite{kamiya}. The formation enthalpies of the generated O-vacancies are estimated to be between 0.77 eV and 4.61 eV(assuming contact with a reservoir of O$_2$ gas). The total energy are affected by the cations around V(O). Generally as the number of In atoms surrounding the V(O) is larger, the formation energy is lowered, and the more stable vacancies tend to have the shallower donor levels, although generally the O-vacancies make donor-like sub-gap levels, as suggested by Kamiya\cite{kamiya}.

Here, we note that the O-deficient a-IGZO is not simply described by the O-vacancy structure. These vacancies are metastable states, thus we tested the additional relaxation of a-IGZO with a missing oxygen atom in a supercell by repeating the SA process. The total energy becomes smaller by 0.45 eV than the most stable V(O) structure. Below, we will discriminate between the metastable V(O) structure and the latter O-deficiency state. The low formation enthalpy (0.32 eV) of the neutral shallow-donor-like O-deficiency state indicates that the (2+)-charged state can be more easily formed, when Fermi level is around mid-gap in this wide-gap materials.

We now examine the structural change by the O-deficiency in a-IGZO, with respect to the short-range order. We find, as shown in Fig. \ref{eff}b, that {\it the $<$CN$^*>$ of In is particularly reduced by the O-deficiency.} It indicates that the O-deficiency should be more easily achieved in the presence of In atoms.The result can be understood by the increase in bond strength as we go from In-O to Zn-O and to Ga-O \cite{bs}, consistent with In having the largest atomic radius and also with the behavior of the stability of V(O). As shown in Fig. \ref{eff}b and c, both $\Delta_{\mathrm{STD}}$CN$^*$ and $\Delta_{\mathrm{STD}}$d$^*$ around In atoms become all increased by O-deficiency, indicating that as the number of O atoms surrounding the large In atoms is reduced, the structural distortions around the In atoms becomes enhanced.

\begin{figure}[b]
\begin{center}
\label{es}
\centering{\includegraphics[width=16.5cm]{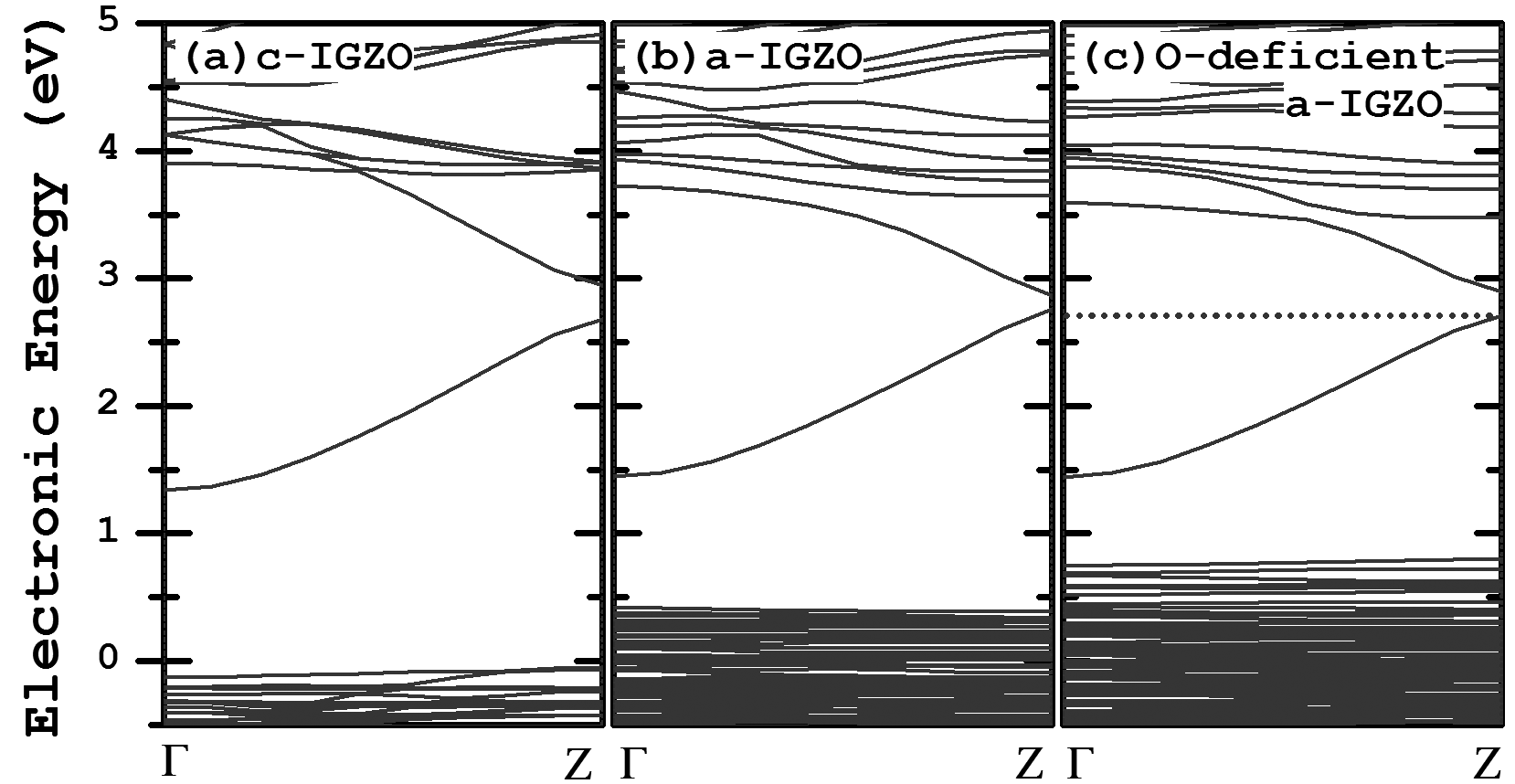}}
\caption{The electronic structure of (a) crystalline c-InGaZnO$_{4}$, (b) stoichiometric amorphous a-InGaZnO$_{4}$
and (c) an O-deficient a-InGaZnO$_{4}$ by one missing oxygen in supercell, where the dotted line describes the
Fermi level. The Brillouin zone is artificial, given by the rhombohedral supercell, and the Z is (0,0, 0.104
\AA$^{-1}$).}
\end{center}
\end{figure}

\begin{figure}
\begin{center}
\label{tipr}
\centering{\includegraphics[width=14.0cm]{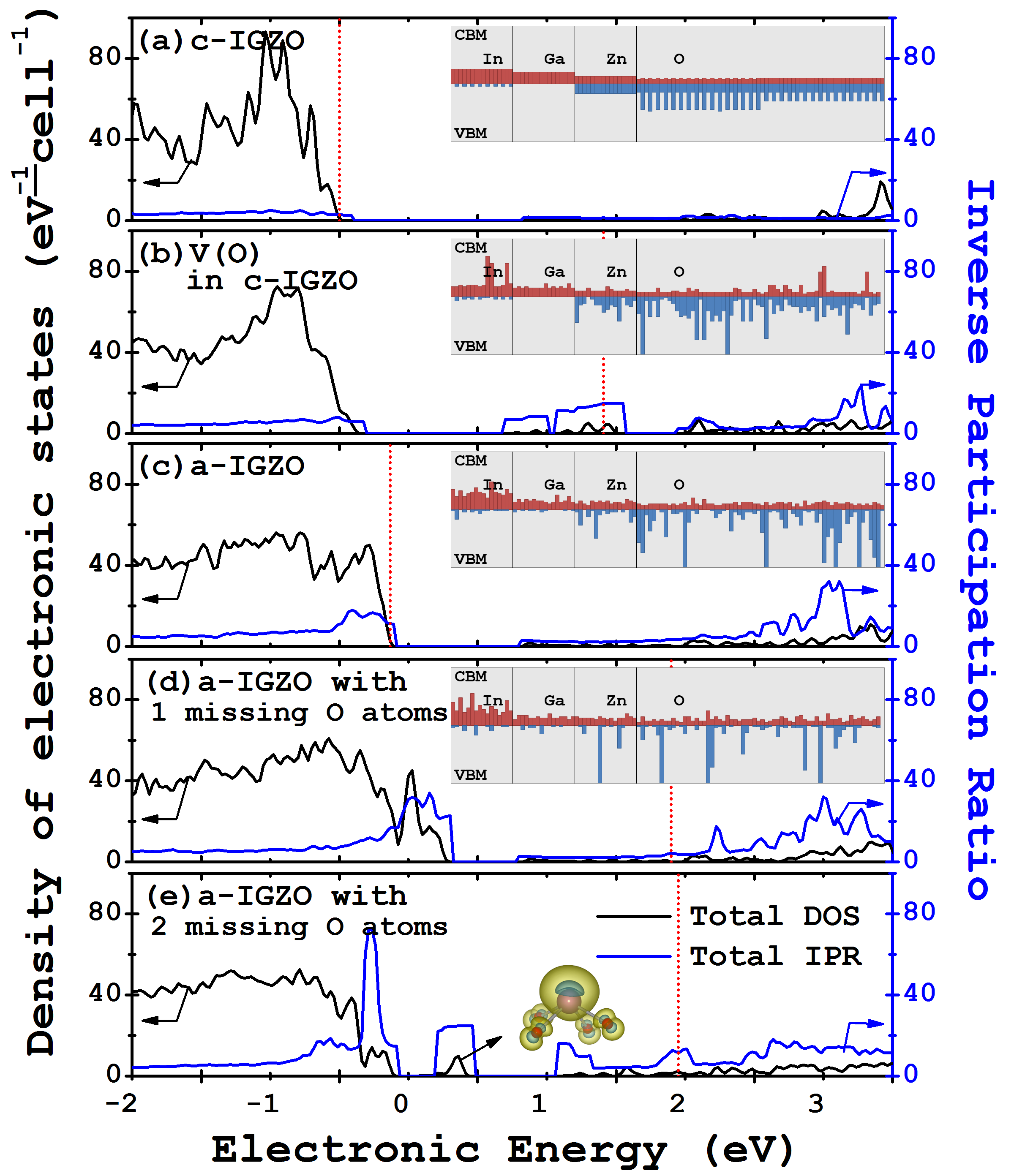}}
\caption{(Color online) The density of electronic states of (a) the c-IGZO and (b) the O-vacancy in c-IGZO, (c)
the stoichiometric a-IGZO, (d) an O-deficient a-IGZO with one missing oxygen atom (e) an O-deficient a-IGZO with
two missing oxygen atoms in supercell are shown. The bar graphs in insets indicate the atomic mappings of the wave
functions at conduction band minimum (brown bars) and valence band maximum (black bars) (see the detail in text)
and the localization of the wave functions are described by the inverse participation ratio (IPR) calculated from
the atomic mapping of the wavefunctions (see the text), and the red dotted lines indicate the Fermi level.}
\end{center}
\end{figure}

The electronic structures of c-IGZO and a-IGZO are compared in Fig. \ref{es}. The localization characters of the wave functions at the conduction band minimum (CBM) and valence band maximum (VBM) are described by the rod graphs shown in the inset of Fig. \ref{tipr}, which show the mapping of wave functions onto atomic sites: we calculated the angular momentum decomposition of a wavefunction: $\Psi = \sum_{i,\alpha} a_{i,\alpha} \phi^i_{\alpha}$, where $i$ denotes atom and $\alpha$ angular momentum, and the atomic site mapping $c_i = \sum_{\alpha} a_{i,\alpha}$. The site-mappings are parameterized by the normalized inverse participation ratio ($\mathrm{IPR} = (1/N \sum{c_i^4})/(1/N \sum {c_i^2})^2$) (see Fig. \ref{tipr}). The IPR increases with the localization and thus can describe the band tail states. The IPR of c-IGZO are usually small.

An important finding is that {\it the fluctuation in the electronic structure induced by the disorder is very small in the CBM.} The character of the electronic structure around the CBM in the disordered a-IGZO is calculated to be amazingly close to that in c-IGZO (see Fig. \ref{es} and \ref{tipr}), accordingly the IPR data are similar to that in c-IGZO. From these results, we suggest that the localized band tail, which is well-known to be formed around the band edges in amorphous materials, can be absent around CBM in the ASMO. Thus the mobility of electron carrier in a-IGZO can be comparable to that in c-IGZO. In the high ionic oxides, the hybridizations between the cation-\emph{s} and O-\emph{p} orbitals are weak. The CBM state is characterized almost only by the cation-$s$-orbitals. Thus the connections between cation-s-orbitals at CBM are insensitive to the CNs of cations, unless the dangling bond is formed. Therefore the character of CBE is insensitive to the disorder.

The wavefunction of the CBM is more accumulated around In atoms in both c-IGZO and a-IGZO (see the inset of Fig. \ref{tipr}a and \ref{tipr}c), indicating that the channel formed among In atoms should be more favorable for the electron transport, thus the higher concentration of In should be favored for the high mobility. Since the CBM wavefunction remains sufficiently well delocalized in the amorphous structure, an effective mass for the free electron carrier can be defined. The calculated effective mass of 0.201 \emph{m$_0$} in a-IGZO is, thus, very close to that in c-IGZO(=0.196 $m_0$ in an ab-plane, and 0.213 $m_0$ along the c-axis).

We note that the effect of the disorder is strong for the valence band. The IPR data indicate that the wavefunctions around the valence band (VB) becomes seriously localized, as shown in Fig. \ref{tipr}c. There is a region around the VBM where the IPR is very large. These indicate, that {\it the localized band tail states are strongly present at the VBM}. It is because the VBM orbital consists of strongly directional O-\emph{p} orbitals. It leads to a lowering of band gap. This should be an important feature of the a-IGZO and the ASMO. The band gap (E$_g$) of a-IGZO is calculated to be smaller by 0.44 eV than that of c-IGZO. The localized tail states should trap hole carriers (not for electron carriers), thus the low mobility of hole carrier should be low. These states can play a role in the illumination stress instability of the TFT \cite{ko,jkyeong}.

We now discuss the effect of O-deficiency (one-O-missing in supercell) on the electronic states, which is compared to that of c-IGZO. The O-deficiency plays a role of donor. Usually the donors creates donor level below the CBM, however surprisingly we could not find any localized states around CBM. The atomic mapping and the IPR value of the wavefunctions around CBM, is little affected by the O-deficiency. These indicate that a localized orbital is not generated by the O-deficiency, \emph{i.e.} in contrast to the case in c-IGZO. The O-deficiency can be described by the change in the substructures around cations rather than by the formation of O-vacancy structure, and just as the CBM is insensitive to the disorder characterized by the admixture of many kinds of substructures, it is not also affected much by the change in substructures. Even if a donor-center is not formed, free electron carriers are generated due to the chemical imbalance induced by the O-deficiency.

On the other hand, in c-IGZO, an O-vacancy gives rise to dangling bonds and create the localized donor-like trap centers. In the c-IGZO which has a layered structure, four kinds of O-vacancy structures can be distinguished. The most stable V(O) is surrounded by three In atoms and one Zn atoms. The V(O) preferentially forms around In atoms \cite{kang}. This trend is consistent with the fact that the O atoms can be easily removed from around In in O-deficient a-IGZO. We find that the formation enthalpies of the O-vacancies in c-IGZO are much larger that in a-IGZO. The formation enthalpies of these O-vacancies are calculated to be 4.14$\sim$4.64 eV. It indicates that O-deficiency can be much more easily achieved in the a-IGZO than in c-IGZO. Even if the electronic structure calculations indicate that the lowest energy V(O) in c-IGZO induces a shallow level, close to CBM, as shown in Fig. \ref{tipr}b, the atomic mapping of the wavefunction indicate that {\it the CBM orbital is slightly localized at In atoms around V(O)}, resulting in the increase of the IPR values. This is in contrast to the case of O-deficient a-IGZO. This result can provide an explanation for how the carrier mobility can be measured to be lower in the c-IGZO rather than in a-IGZO, unless the O-deficiency is too serious \cite{nomura2}. The mobility can increase with the carrier concentration by the increasing O-deficiency concentration\cite{nomura2}, since the furthermore formation of trap centers can be avoided by the O-deficiency in a-IGZO\cite{kamiya2}.

For the valence band, the effect of the O-deficiency is to create the localized states more above the VBM. Accordingly the calculated band gap (E$_g$) was more lowered by 0.59 eV, compared to the stoichiometric case, as shown by Fig. \ref{es}c. These states can provide an explanation for the absorption of blue-green light. {\it As the O-deficiency concentration becomes more serious by two-more missing-O atoms per supercell, the structural fluctuation becomes more pronounced and more complex, and sub-structures with weak or dangling bonds are formed}. Some d$^*$ are calculated to be much larger than the fitting line for the a-IGZO with the two-more missing atoms, as shown in Fig. \ref{eff}a. Both $\Delta_{\mathrm{STD}}$ d$^*$ and $\Delta_{\mathrm{STD}}$CN$^*$ becomes distinctly larger around In atoms by the increased O-deficiency(see Fig. \ref{eff}), indicating that the local structures around In atoms become remarkably distorted and thus weak or dangling bond tend to be generated around In. In the case of two missing oxygens, we identify an asymmetric InO$_4$ structure having an isolated dangling bond (see the electron density picture in Fig. \ref{tipr}e). It induces a deep donor level within gap, trapping two electrons. Furthermore the conduction band minimum is also serious altered, that should affect the electron mobility. We repeated several SA processes, but could not avoid the deep level formation. The reduction of the electronic energy resulting from the capture of electrons at the deep level may stabilize this kind of DB formation. This kind of deep level may lead to the Fermi level pinning in the seriously O-deficient a-IGZO, and may give an explanation for the subgap states\cite{kamiya2}, and also reduce the mobility. We note that the subgap states can also come from the metastable vacancy structures.

In conclusion, we investigated the microscopic and electronic structures of a-IGZO as a representative of ASMO. We showed that the band tail states, which are well-known to be present in amorphous semiconductor, can be absent around the CBM in amorphous IGZO, even if the structure is complicated by the admixture of many different kinds of substructures. The properties of the CBM are found to be little affected even by the donor-like O-deficiency, which is in contrast to the crystalline IGZO, where an O-vacancy forms a localized state. These give an explanation to why the mobility of amorphous IGZO can be much higher than that of crystalline IGZO. The disorder and O-deficiency are found to affect mainly the character of the valence band, resulting in the serious formation of localized
states around the VBM. Our results provide a new perspective on the physics of the ASMO.

\begin{acknowledgments}
This work was supported by the National Research Foundation of Korea(NRF) grant funded by the Korea
government(MEST) (No. 2011-0001566). We thank Dr. D. J. Chadi for careful reading and comments.
\end{acknowledgments}

\bibliographystyle{apsrev4-1}

\begin{thebibliography}{16}%
\makeatletter
\providecommand \@ifxundefined [1]{%
 \@ifx{#1\undefined}
}%
\providecommand \@ifnum [1]{%
 \ifnum #1\expandafter \@firstoftwo
 \else \expandafter \@secondoftwo
 \fi
}%
\providecommand \@ifx [1]{%
 \ifx #1\expandafter \@firstoftwo
 \else \expandafter \@secondoftwo
 \fi
}%
\providecommand \natexlab [1]{#1}%
\providecommand \enquote  [1]{``#1''}%
\providecommand \bibnamefont  [1]{#1}%
\providecommand \bibfnamefont [1]{#1}%
\providecommand \citenamefont [1]{#1}%
\providecommand \href@noop [0]{\@secondoftwo}%
\providecommand \href [0]{\begingroup \@sanitize@url \@href}%
\providecommand \@href[1]{\@@startlink{#1}\@@href}%
\providecommand \@@href[1]{\endgroup#1\@@endlink}%
\providecommand \@sanitize@url [0]{\catcode `\\12\catcode `\$12\catcode
  `\&12\catcode `\#12\catcode `\^12\catcode `\_12\catcode `\%12\relax}%
\providecommand \@@startlink[1]{}%
\providecommand \@@endlink[0]{}%
\providecommand \url  [0]{\begingroup\@sanitize@url \@url }%
\providecommand \@url [1]{\endgroup\@href {#1}{\urlprefix }}%
\providecommand \urlprefix  [0]{URL }%
\providecommand \Eprint [0]{\href }%
\providecommand \doibase [0]{http://dx.doi.org/}%
\providecommand \selectlanguage [0]{\@gobble}%
\providecommand \bibinfo  [0]{\@secondoftwo}%
\providecommand \bibfield  [0]{\@secondoftwo}%
\providecommand \translation [1]{[#1]}%
\providecommand \BibitemOpen [0]{}%
\providecommand \bibitemStop [0]{}%
\providecommand \bibitemNoStop [0]{.\EOS\space}%
\providecommand \EOS [0]{\spacefactor3000\relax}%
\providecommand \BibitemShut  [1]{\csname bibitem#1\endcsname}%
\let\auto@bib@innerbib\@empty
\bibitem [{\citenamefont {Nomura}\ \emph {et~al.}(2003)\citenamefont {Nomura}
  \emph {et~al.}}]{nomura1}%
  \BibitemOpen
  \bibfield  {author} {\bibinfo {author} {\bibfnamefont {K.}~\bibnamefont
  {Nomura}} \emph {et~al.},\ }\href@noop {} {\bibfield  {journal} {\bibinfo
  {journal} {Science}\ }\textbf {\bibinfo {volume} {300}},\ \bibinfo {pages}
  {1269} (\bibinfo {year} {2003})}\BibitemShut {NoStop}%
\bibitem [{\citenamefont {Kamiya}\ and\ \citenamefont
  {Hosono}(2010)}]{kamiya2}%
  \BibitemOpen
  \bibfield  {author} {\bibinfo {author} {\bibfnamefont {T.}~\bibnamefont
  {Kamiya}}\ and\ \bibinfo {author} {\bibfnamefont {H.}~\bibnamefont
  {Hosono}},\ }\href@noop {} {\bibfield  {journal} {\bibinfo  {journal} {NPG
  Asia Mater.}\ }\textbf {\bibinfo {volume} {2}},\ \bibinfo {pages} {1522}
  (\bibinfo {year} {2010})}\BibitemShut {NoStop}%
\bibitem [{\citenamefont {Jeong}(2011)}]{jkyeong}%
  \BibitemOpen
  \bibfield  {author} {\bibinfo {author} {\bibfnamefont {J.~K.}\ \bibnamefont
  {Jeong}},\ }\href@noop {} {\bibfield  {journal} {\bibinfo  {journal}
  {Semicond. Sci. Techol.}\ }\textbf {\bibinfo {volume} {26}},\ \bibinfo
  {pages} {034008} (\bibinfo {year} {2011})}\BibitemShut {NoStop}%
\bibitem [{\citenamefont {Nomura}\ \emph {et~al.}(2004)\citenamefont {Nomura}
  \emph {et~al.}}]{nomura2}%
  \BibitemOpen
  \bibfield  {author} {\bibinfo {author} {\bibfnamefont {K.}~\bibnamefont
  {Nomura}} \emph {et~al.},\ }\href@noop {} {\bibfield  {journal} {\bibinfo
  {journal} {Nature (London)}\ }\textbf {\bibinfo {volume} {488}},\ \bibinfo
  {pages} {432} (\bibinfo {year} {2004})}\BibitemShut {NoStop}%
\bibitem [{\citenamefont {Yao}\ \emph {et~al.}(2011)\citenamefont {Yao} \emph
  {et~al.}}]{jianke}%
  \BibitemOpen
  \bibfield  {author} {\bibinfo {author} {\bibfnamefont {J.}~\bibnamefont
  {Yao}} \emph {et~al.},\ }\href@noop {} {\bibfield  {journal} {\bibinfo
  {journal} {IEEE Trans. Electron Devices}\ }\textbf {\bibinfo {volume} {58}},\
  \bibinfo {pages} {1121} (\bibinfo {year} {2011})}\BibitemShut {NoStop}%
\bibitem [{\citenamefont {Kamiya}\ \emph {et~al.}(2008)\citenamefont {Kamiya}
  \emph {et~al.}}]{kamiya}%
  \BibitemOpen
  \bibfield  {author} {\bibinfo {author} {\bibfnamefont {T.}~\bibnamefont
  {Kamiya}} \emph {et~al.},\ }\href@noop {} {\bibfield  {journal} {\bibinfo
  {journal} {Phys. Stat. Sol. (c)}\ }\textbf {\bibinfo {volume} {5}},\ \bibinfo
  {pages} {3098} (\bibinfo {year} {2008})}\BibitemShut {NoStop}%
\bibitem [{\citenamefont {Shin}\ \emph {et~al.}(2009)\citenamefont {Shin} \emph
  {et~al.}}]{ko}%
  \BibitemOpen
  \bibfield  {author} {\bibinfo {author} {\bibfnamefont {J.~H.}\ \bibnamefont
  {Shin}} \emph {et~al.},\ }\href@noop {} {\bibfield  {journal} {\bibinfo
  {journal} {ETRI J.}\ }\textbf {\bibinfo {volume} {31}},\ \bibinfo {pages}
  {62} (\bibinfo {year} {2009})}\BibitemShut {NoStop}%
\bibitem [{\citenamefont {Bl$\ddot{\mathrm{o}}$chl}(1994)}]{paw}%
  \BibitemOpen
  \bibfield  {author} {\bibinfo {author} {\bibfnamefont {P.~E.}\ \bibnamefont
  {Bl$\ddot{\mathrm{o}}$chl}},\ }\href@noop {} {\bibfield  {journal} {\bibinfo
  {journal} {Phys. Rev. B}\ }\textbf {\bibinfo {volume} {50}},\ \bibinfo
  {pages} {17953} (\bibinfo {year} {1994})}\BibitemShut {NoStop}%
\bibitem [{\citenamefont {Kresse}\ and\ \citenamefont
  {Furthmuller}(1996)}]{vasp}%
  \BibitemOpen
  \bibfield  {author} {\bibinfo {author} {\bibfnamefont {G.}~\bibnamefont
  {Kresse}}\ and\ \bibinfo {author} {\bibfnamefont {J.}~\bibnamefont
  {Furthmuller}},\ }\href@noop {} {\bibfield  {journal} {\bibinfo  {journal}
  {Phys. Rev. B}\ }\textbf {\bibinfo {volume} {54}},\ \bibinfo {pages} {11169}
  (\bibinfo {year} {1996})}\BibitemShut {NoStop}%
\bibitem [{\citenamefont {Perdew}\ \emph {et~al.}(1996)\citenamefont {Perdew},
  \citenamefont {Burke},\ and\ \citenamefont {Ernzerhof}}]{pbe}%
  \BibitemOpen
  \bibfield  {author} {\bibinfo {author} {\bibfnamefont {J.~P.}\ \bibnamefont
  {Perdew}}, \bibinfo {author} {\bibfnamefont {K.}~\bibnamefont {Burke}}, \
  and\ \bibinfo {author} {\bibfnamefont {M.}~\bibnamefont {Ernzerhof}},\
  }\href@noop {} {\bibfield  {journal} {\bibinfo  {journal} {Phys. Rev. Lett.}\
  }\textbf {\bibinfo {volume} {77}},\ \bibinfo {pages} {3865} (\bibinfo {year}
  {1996})}\BibitemShut {NoStop}%
\bibitem [{\citenamefont {Anisimov}\ \emph {et~al.}(1991)\citenamefont
  {Anisimov}, \citenamefont {Zaanen},\ and\ \citenamefont {Andersen}}]{ldau}%
  \BibitemOpen
  \bibfield  {author} {\bibinfo {author} {\bibfnamefont {V.~I.}\ \bibnamefont
  {Anisimov}}, \bibinfo {author} {\bibfnamefont {J.}~\bibnamefont {Zaanen}}, \
  and\ \bibinfo {author} {\bibfnamefont {O.~K.}\ \bibnamefont {Andersen}},\
  }\href@noop {} {\bibfield  {journal} {\bibinfo  {journal} {Phys. Rev. B}\
  }\textbf {\bibinfo {volume} {44}},\ \bibinfo {pages} {943} (\bibinfo {year}
  {1991})}\BibitemShut {NoStop}%
\bibitem [{\citenamefont {Hoppe}(1970)}]{effcn1}%
  \BibitemOpen
  \bibfield  {author} {\bibinfo {author} {\bibfnamefont {R.}~\bibnamefont
  {Hoppe}},\ }\href@noop {} {\bibfield  {journal} {\bibinfo  {journal} {Angew.
  Chem. Int. Ed. Engl.}\ }\textbf {\bibinfo {volume} {9}},\ \bibinfo {pages}
  {25} (\bibinfo {year} {1970})}\BibitemShut {NoStop}%
\bibitem [{\citenamefont {Hoppe}(1979)}]{effcn2}%
  \BibitemOpen
  \bibfield  {author} {\bibinfo {author} {\bibfnamefont {R.}~\bibnamefont
  {Hoppe}},\ }\href@noop {} {\bibfield  {journal} {\bibinfo  {journal} {Z.
  Kristallogr.}\ }\textbf {\bibinfo {volume} {150}},\ \bibinfo {pages} {23}
  (\bibinfo {year} {1979})}\BibitemShut {NoStop}%
\bibitem [{\citenamefont {Lide}(1996)}]{bs}%
  \BibitemOpen
  \bibfield  {author} {\bibinfo {author} {\bibfnamefont {D.~L.}\ \bibnamefont
  {Lide}},\ }\href@noop {} {\emph {\bibinfo {title} {CRC Handbook of Chemistry
  and Physics}}},\ \bibinfo {edition} {77th}\ ed.\ (\bibinfo  {publisher} {CRC
  press, Boca Raton},\ \bibinfo {address} {Boca Raton},\ \bibinfo {year}
  {1996})\BibitemShut {NoStop}%
\bibitem [{\citenamefont {Kang}\ and\ \citenamefont {Park}(2010)}]{kang}%
  \BibitemOpen
  \bibfield  {author} {\bibinfo {author} {\bibfnamefont {I.~J.}\ \bibnamefont
  {Kang}}\ and\ \bibinfo {author} {\bibfnamefont {C.~H.}\ \bibnamefont
  {Park}},\ }\href@noop {} {\bibfield  {journal} {\bibinfo  {journal} {J.
  Korean Phys. Soc.}\ }\textbf {\bibinfo {volume} {56}},\ \bibinfo {pages}
  {480} (\bibinfo {year} {2010})}\BibitemShut {NoStop}%
\end{thebibliography}

\providecommand{\noopsort}[1]{}\providecommand{\singleletter}[1]{#1}%

\end{document}